\documentclass[12pt,a4paper]{article}
\usepackage{amsmath}
\usepackage[font={footnotesize,it}]{caption}
\usepackage[top=1.4in, bottom=1.4in, left=1.3in, right=1.3in]{geometry}

\DeclareCaptionStyle{italic}[justification=centering]{labelfont={bf},textfont={it},labelsep=colon}
\usepackage{graphicx,psfrag,epsf}
\usepackage{enumerate}
\usepackage{natbib}
\usepackage[table]{xcolor}
\usepackage{url, setspace}
\usepackage{booktabs, subfig, bm, paralist,mathpazo,tikz,todonotes,longtable,microtype,dsfont,rotating} 
\usepackage[pdftex,colorlinks=true]{hyperref}
\definecolor{darkblue}{rgb}{0,0,.6}
\hypersetup{citecolor=darkblue,linkcolor=darkblue,urlcolor=darkblue}

\newcommand{\blind}{0}

\newcommand{\E}{\text{E}}
\newcommand{\F}{\mathcal{F}}
\newcommand{\G}{\mathcal{G}}
\newcommand{\Var}{\text{Var}}

\newcommand{\X}{\mathcal{X}}
\newcommand{\Y}{\mathcal{Y}}
\graphicspath{{plots/}}

\newsavebox\CBox

\date{\today}

\begin{document}

\def\spacingset#1{\renewcommand{\baselinestretch}
{#1}\small\normalsize} \spacingset{1}

\if0\blind
{
  \title{\bf Granger causality of bivariate stationary curve time series}
  \author{Han Lin Shang\thanks{Postal address: Department of Actuarial Studies and Business Analytics, Level 7, 4 Eastern Road, Macquarie University, Sydney, NSW 2109, Australia; Telephone number: +61(2) 9850 4689; Email: hanlin.shang@mq.edu.au; ORCID: \url{https://orcid.org/0000-0003-1769-6430}.}
  \hspace{.2cm}\\
    Department of Actuarial Studies and Business Analytics \\
    Macquarie University\\
\\
    Kaiying Ji \\
    Discipline of Accounting \\
    The University of Sydney \\    
\\
Ufuk Beyaztas \\
Department of Mathematics \\
Bartin University
}
  \maketitle
} \fi

\if1\blind
{
   \title{\bf Granger causality of bivariate stationary curve time series}
   \author{}
   \maketitle
} \fi

\bigskip

\begin{abstract}
We study causality between bivariate curve time series using the Granger causality generalized measures of correlation. With this measure, we can investigate which curve time series Granger-causes the other; in turn, it helps determine the predictability of any two curve time series. Illustrated by a climatology example, we find that the sea surface temperature Granger-causes the sea-level atmospheric pressure. Motivated by a portfolio management application in finance, we single out those stocks that lead or lag behind Dow-Jones industrial averages. Given a close relationship between S\&P 500 index and crude oil price, we determine the leading and lagging variables.
\\

\noindent \textit{Keywords}: Granger causality; G-causality; functional time series.
\end{abstract}

\newpage
\spacingset{1.58}

\section{Introduction}

In recent years, there has been increasing interest in studying the time series of functions observed at high frequency. In these time series, the data frequency is high enough to model itself as a curve time series and gives rise to the analysis of functional time series \citep[see, e.g.,][]{HK12, KR17}. Examples of functional time series include intraday stock price curves with each functional observation defined as a pricing function of time points within a day \citep[e.g., see][]{HKR14, LRS20}, and intraday volatility curves with each functional observation defined as a volatility function of time points within a day \citep[e.g., see][]{SYK19}.

Most of the existing literature focuses on statistical inference, modeling, and forecasting of a univariate functional time series. We study co-movement by observing samples of bivariate stationary curve time series. A natural question is which time series causes the other? In other words, which functional variable is the leading variable and which functional variable is the lagging variable. To address this problem, we use a measure of causality in scalar-valued univariate time series analysis, known as Granger causality: Based on a suggestion by \cite{Wiener56}, \cite{Granger69} proposed a Granger causality measure that relies on the \textit{additional} predictive ability of the second variable. Granger causality has since been extended and applied to a range of research fields. \cite{Granger80} considered testing for causality, while \cite{Granger88} drew a connection between causation (with a lag between cause and effect) and co-integration. \cite{HJ94} tested linear and non-linear Granger causality in the stock price-volume relation, while \cite{DP06} introduced a nonparametric test for Granger non-causality. In neuroscience and neuroimaging, \cite{SBB15} applied Granger causality to study neural activity using electrophysiological and fMRI data.

The Granger causality measure has been modified to generalized measures of correlation (GMC) and Granger causality generalized measures of correlation (GcGMC) by \cite{ZSZ12}. \cite{Vinod17} applied the GMC criterion to analyze the development of economic markets in a study of 198 countries. Further, \cite{Vinod19} developed a package in R \citep{Team20} called ``generalCorr" to implement the GMC. \cite{AH18} used the GMC criterion to analyze causal relations between the VIX, S\&P 500, and the realized volatility of the S\&P 500 sampled at five-minute intervals. \cite{CLC+17} extended the GMC criterion to propose a model-free feature screening approach, namely sure explained variability and independence screening. In this paper, we aim to extend GcGMC from bivariate scalar-valued time series to bivariate function-valued time series.

The outline of this paper is as follows. In Section~\ref{sec:1_2}, we present a background for GMC and GcGMC. In Section~\ref{sec:2}, we present a nonparametric function-on-function regression model to estimate conditional mean in function-valued GcGMC. Through a series of simulation studies in Section~\ref{sec:3}, we examine the finite-sample performance of our estimate. In Section~\ref{sec:4}, we present three data analyses for our motivating data examples in climatology and finance, respectively. Conclusions are given in Section~\ref{sec:5}.

\section{Granger causality generalized measures of correlation}\label{sec:1_2}

Before introducing GcGMC, we revisit GMC and auto generalized measures of correlation (AGMC). The GMC can be derived from a well-known \textit{variance decomposition} formula,
\begin{equation}
\Var(X) = \Var[\E(X|Y)] + \E[\Var(X|Y)],\label{eq:1}
\end{equation}
where $X$ and $Y$ are second-order stationary processes with finite variances, and $\Var[\E(X|Y)]$ denotes the variance of the conditional mean of $X$ given $Y$. Equation~\eqref{eq:1} states that the unconditional variance of a random variable $X$ can be expressed as the variance of conditional mean plus the expectation of conditional variance.

By dividing~\eqref{eq:1} by Var$(X)$, we obtain
\begin{align*}
1 &= \frac{\Var[\E(X|Y)]}{\Var(X)} + \frac{\E[\Var(X|Y)]}{\Var(X)}, \\
\frac{\Var[\E(X|Y)]}{\Var(X)} &= 1 - \frac{\E[\Var(X|Y)]}{\Var(X)},
\end{align*}
where $\text{GMC}(X|Y) = \frac{\Var[\E(X|Y)]}{\Var(X)}$ can be interpreted as explained variance of $X$ by $Y$. Similarly, we can define GMC$(Y|X)$ as
\begin{equation*}
\text{GMC}(Y|X) = \frac{\Var[\E(Y|X)]}{\Var(Y)} =  1 - \frac{\E[\Var(Y|X)]}{\Var(Y)}.
\end{equation*}
When one models the relationship between random variables $X$ and $Y$ by a linear model $Y = g(X)+\varepsilon$, then GMC$(Y|X)$ is identical to a functional version of $R^2$ when $g(X)$ corresponds to $\E(Y|X)$. However, the GMC is a more generalized correlation measure than the $R^2$, as it can measure possible nonlinear association.

When $Y$ is a lagged variable of $X$, and they are bivariate stationary time series, we can measure their serial auto-correlation by AGMC. It is defined as
\begin{equation*}
\text{AGMC}_k(X_t) = \text{GMC}(X_t|X_{t-k}),
\end{equation*}
where $k>0$ denotes a lag variable. When $(X_t, Y_t)_{t=1,\dots,n}$ are a pair of bivariate scalar time series, we can also measure cross-correlation by AGMC defined as
\begin{equation*}
\text{AGMC}_k(Y_t|X_t) = \text{GMC}(Y_t|X_{t-k}).
\end{equation*}

By taking into account the cross-correlation and auto-correlation, \cite{ZSZ12} proposed a Granger causality general measure of correlation (GcGMC). It is defined as
\begin{align*}
\text{GcGMC}(X_t|\F_{t-1}) &= 1 - \frac{\E[\{X_t - \E(X_t|X_{t-1}, X_{t-2},\dots, Y_{t-1}, Y_{t-2},\dots)\}^2]}{\E(\Var(X_t|X_{t-1}, X_{t-2},\dots))} \\
&= 1 - \frac{\E[\{X_t - \E(X_t|X_{t-1}, X_{t-2},\dots, Y_{t-1}, Y_{t-2},\dots)\}^2]}{\E[\{X_t - \E(X_t|X_{t-1}, X_{t-2},\dots)\}^2]} \\
\text{GcGMC}(Y_t|\F_{t-1}) &= 1 - \frac{\E[\{Y_t - \E(Y_t|Y_{t-1}, Y_{t-2},\dots, X_{t-1}, X_{t-2},\dots)\}^2]}{\E(\Var(Y_t|Y_{t-1}, Y_{t-2},\dots))} \\
&= 1 - \frac{\E[\{Y_t - \E(Y_t|Y_{t-1}, Y_{t-2},\dots, X_{t-1}, X_{t-2},\dots)\}^2]}{\E[\{Y_t - \E(Y_t|Y_{t-1}, Y_{t-2},\dots)\}^2]},
\end{align*}
where $\F_{t-1} = \sigma(X_{t-1}, X_{t-2}, \dots, Y_{t-1}, Y_{t-2},\dots)$ denotes all available series up to time point $t-1$ for a function $\sigma$. 

Suppose that $\{\X(u), \Y(v)\}$ are second-order stationary curve time series, where $u\in [a, b]$ and $v\in [c, d]$ denote two function supports. Those function support ranges can be different. The GcGMC can be expressed as
\begin{align}
\text{GcGMC}(\X_t(u)|\G_{t-1}) &= 1 - \frac{\E[\{\X_t(u) - \E(\X_t(u)|\X_{t-1}(u), \X_{t-2}(u),\dots, \Y_{t-1}(v), \Y_{t-2}(v),\dots)\}^2]}{\E[\{\X_t(u) - \E(\X_t(u)|\X_{t-1}(u),\X_{t-2}(u),\dots)\}^2]} \label{eq:7}\\
\text{GcGMC}(\Y_t(v)|\G_{t-1}) &= 1 - \frac{\E[\{\Y_t(v) - \E(\Y_t(v)|\Y_{t-1}(v), \Y_{t-2}(v), \dots, \X_{t-1}(u), \X_{t-2}(u),\dots)\}^2]}{\E[\{\Y_t(v) - \E(\Y_t(v)|\Y_{t-1}(v),\Y_{t-2}(v),\dots)\}^2]},\label{eq:8}
\end{align}
where $\G_{t-1} = \sigma[\X_{t-1}(u), \X_{t-2}(u),\dots,\Y_{t-1}(v),\Y_{t-2}(v),\dots]$ denotes all available series up to time point $t-1$ for a function $\sigma$.

\section{Function-on-function regression}\label{sec:2}

\subsection{Functional time series}

Functional time series consist of random functions observed at regular time intervals. Functional time series can be classified into two categories depending on if the continuum is also a time variable. On the one hand, functional time series can arise from measurements obtained by separating an almost continuous time record into consecutive intervals \citep[e.g., days or years, see][]{HK12}. We refer to such data structure as sliced functional time series, examples of which include intraday stock price curves \citep{KRS17} and intraday particulate matter \citep{Shang17}. On the other hand, when the continuum is not a time variable, functional time series can also arise when observations over a period are considered as finite-dimensional realizations of an underlying continuous function \citep[e.g., yearly age-specific mortality rates, see][]{LRS20}.

\subsection{Nonparametric function-on-function regression}

Let $\bm{\Y}=(\Y_1, \Y_2,\dots, \Y_n)^{\top}$ be a vector of functional responses and $\bm{\X}=(\X_1, \X_2, \dots, \X_n)^{\top}$ be a vector of functional predictors. Through samples of ($\bm{\X}, \bm{\Y}$), we investigate the causality between bivariate functional time series. We assume ($\bm{\X}, \bm{\Y}$) are second-order stationary. We consider a function-on-function regression with homoskedastic errors. Given observations $(\X_t, \Y_t)_{t=1,2,\dots,n}$, the regression model can be expressed as
\begin{equation*}
\Y_t = m(\X_t)+\varepsilon_t,
\end{equation*}
where $m(\cdot)$ is a smooth function from square-integrable function space to square-integrable function space, and $\varepsilon_t$ denotes error function. When $\X_t=\Y_{t-1}$, we can also model first-order autocorrelation of series $\Y$ by a nonparametric function-on-function regression. Similarly, when $\Y_t = \X_{t+1}$, we can also model first-order autocorrelation of series $\X$ by the nonparametric function-on-function regression.

While functional linear regression can measure a linear association between functional predictor and response, it is more usual to consider a possible nonlinear association between two functional variables. There is an increasing amount of literature on the development of nonparametric functional estimators, such as the functional Nadaraya-Watson (NW) estimator \citep{FV06}. For estimating the conditional mean, the functional NW estimator can be defined as
\begin{equation}
\widehat{m}_h(\X, \bm{\Y}) = \sum^n_{t=1}\frac{K_h[d(\X_t, \X)]}{\sum^n_{t=1}K_h[d(\X_t, \X)]}\Y_t, \label{eq:9}\\
\end{equation}
where $K(\cdot)$ denotes a kernel function that integrates to one. It is often chosen as a unimodal probability density function that can be either symmetric or non-symmetric around zero and has a finite variance. Here, we choose the quadratic kernel function.

It is often assumed that functions have a continuous derivative on the function support range for measuring the distance between two curves. We compute a semi-metric $d(\cdot, \cdot)$ based on a second-order derivative, given by
\begin{equation*}
d_2^{\text{deriv}}(\X_t, \X) = \sqrt{\int_a^b \Big[\X_t^{(2)}(u) - \X^{(2)}(u)\Big]^2du},
\end{equation*}
where $\X^{(2)}(u)$ is the second-order derivative of $\X(u)$.

The bandwidth parameter $h$ controls the trade-off between squared bias and variance in the mean squared error given by $\E[m(\X) - \widehat{m}(\X)]^2$, where $\widehat{m}(\X)$ is an estimator of the true but unknown regression function $m(\X)$. Here, we choose $h$ by generalized cross validation.

Without knowing it as a priori, the first-order temporal dependence is often adequate and convenient to model a time series \citep[see also][]{Granger88, TBE14}. When the functional response variable is one-lag-ahead of the functional predictor variable, we can use the functional NW estimator in~\eqref{eq:9} to capture the autocorrelation. It can be expressed as:
\begin{equation}
\widehat{m}_b(\X) = \sum^{n-1}_{t=1}\frac{K_b[d(\X_t, \X)]}{\sum^{n-1}_{t=1}K_b[d(\X_t, \X)]}\X_{t+1}. \label{eq:10b}
\end{equation}
With the estimated bandwidth parameter $b$ and functional NW estimator in~\eqref{eq:10b}, we can obtain a one-step-ahead prediction of $\X_{n+1}$ given by
\begin{equation*}
\widehat{\X}_{n+1} = \widehat{m}_b(\X_n) = \sum^{n-1}_{t=1}\frac{K_h[d(\X_t, \X_n)]}{\sum^{n-1}_{t=1}K_h[d(\X_t, \X_n)]}\X_{t+1}.
\end{equation*}
By plugging estimated $\widehat{\X}_{n+1}$ into~\eqref{eq:9}, we obtain a one-step-ahead prediction of $\Y_{n+1}$ given by
\begin{equation*}
\widehat{\Y}_{n+1} = \widehat{m}_h(\widehat{\X}_{n+1}, \bm{\Y}) = \sum^n_{t=1}\frac{K_h[d(\X_t, \widehat{\X}_{n+1})]}{\sum^n_{t=1}K_h[d(\X_t, \widehat{\X}_{n+1})]}\Y_t.
\end{equation*}
With the forecast and holdout functions, we can compute the one-step-ahead prediction error $\mathcal{E}_{n+1}(\Y) = \Y_{n+1} - \widehat{\Y}_{n+1}$. 

\subsection{GcGMC for curve time series}

The GcGMC defined in~\eqref{eq:7} and~\eqref{eq:8} can be viewed as a prediction problem. We divide our data set into a training sample consisting of a portion of the data and a testing sample consisting of the remaining data. For the $t$\textsuperscript{th} observation in the testing sample, the ratio of mean square prediction error can be expressed as:
\begin{align}
\text{GcGMC}(\X) &= 1- \frac{\E\{[\X_{t}(u) - \E(\X_{t}(u)|\X_{t-1}(u),\X_{t-2}(u),\dots,\Y_{t-1}(v),\Y_{t-2}(v),\dots)]^2\}}{\E\{[\X_{t}(u) - \E(\X_{t}(u)|\X_{t-1}(u),\X_{t-2}(u),\dots)]^2\}}  \label{eq:10}\\
\text{GcGMC}(\Y) &= 1 - \frac{\E\{[\Y_{t}(v) - \E(\Y_{t}(v)|\Y_{t-1}(v),\Y_{t-2}(v),\dots,\X_{t-1}(u),\X_{t-2}(u),\dots)]^2\}}{\E\{[\Y_{t}(v) - \E(\Y_{t}(v)|\Y_{t-1}(v),\Y_{t-2}(v),\dots)]^2\}}, \label{eq:11}
\end{align}
where $\E(\cdot)$ denotes conditional expectation.

The GcGMC$(\X)$ and GcGMC$(\Y)$ provide an overall measure of the GcGMC$(\X_t(u)|\mathcal{G}_{t-1})$ and GcGMC$(\Y_t(v)|\mathcal{G}_{t-1})$ in~\eqref{eq:7} and~\eqref{eq:8}. The numerator and denominator in~\eqref{eq:10} and~\eqref{eq:11} measure sum squared prediction errors. If $\Y$ Granger-causes $\X$, the inclusion of $\Y$ information can help to reduce the sum squared prediction errors in $\X$ \citep[see also][]{GT87}. Thus, it results in a positive-valued GcGMC. Similarly, if $\X$ Granger-causes $\Y$, the addition of $\X$ information can help to reduce the sum squared prediction errors in $\Y$.

\section{Simulation studies}\label{sec:3}

\begin{figure}[!htbp]
\centering
{\includegraphics[width=8.2cm]{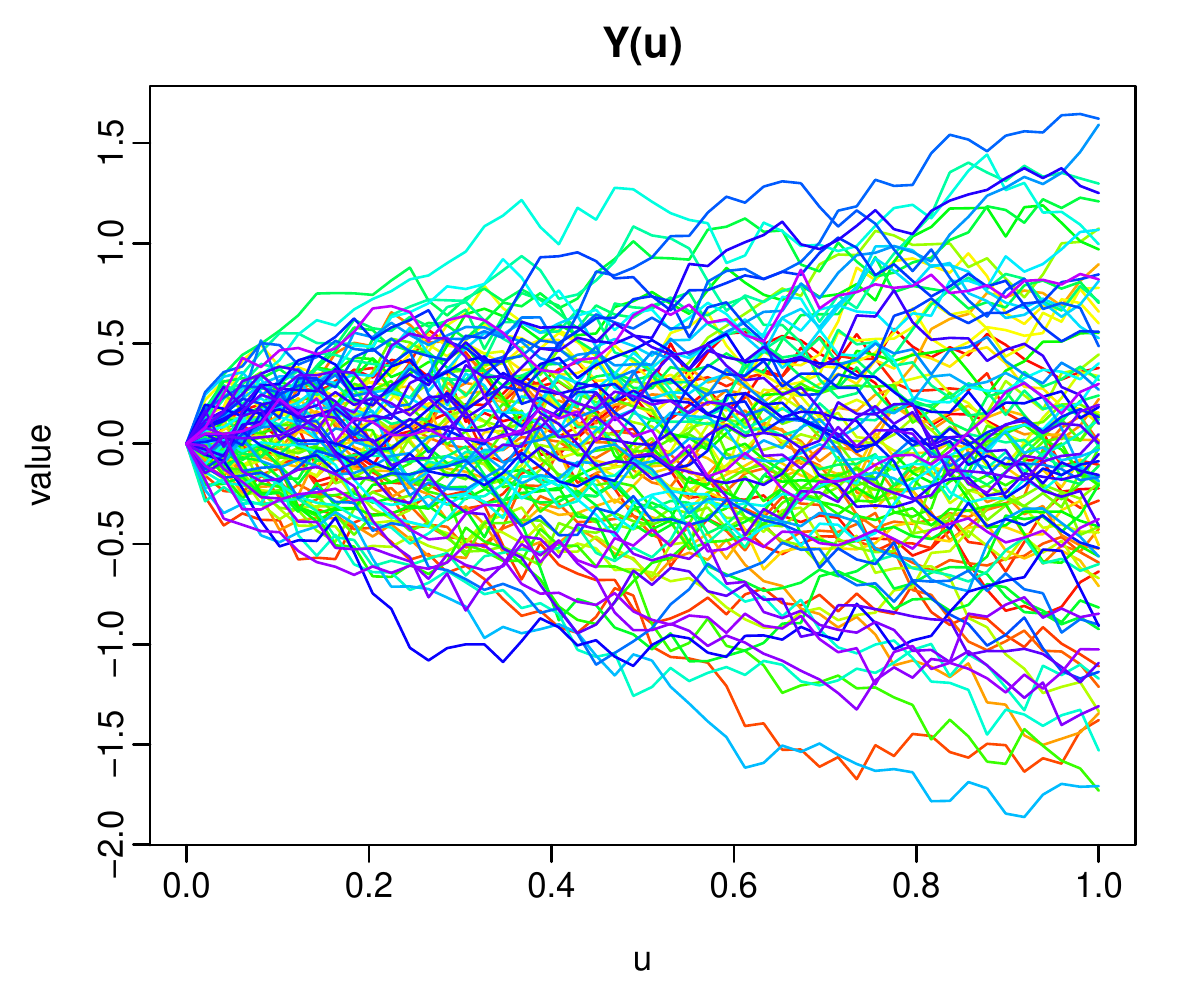}}
\quad
{\includegraphics[width=8.2cm]{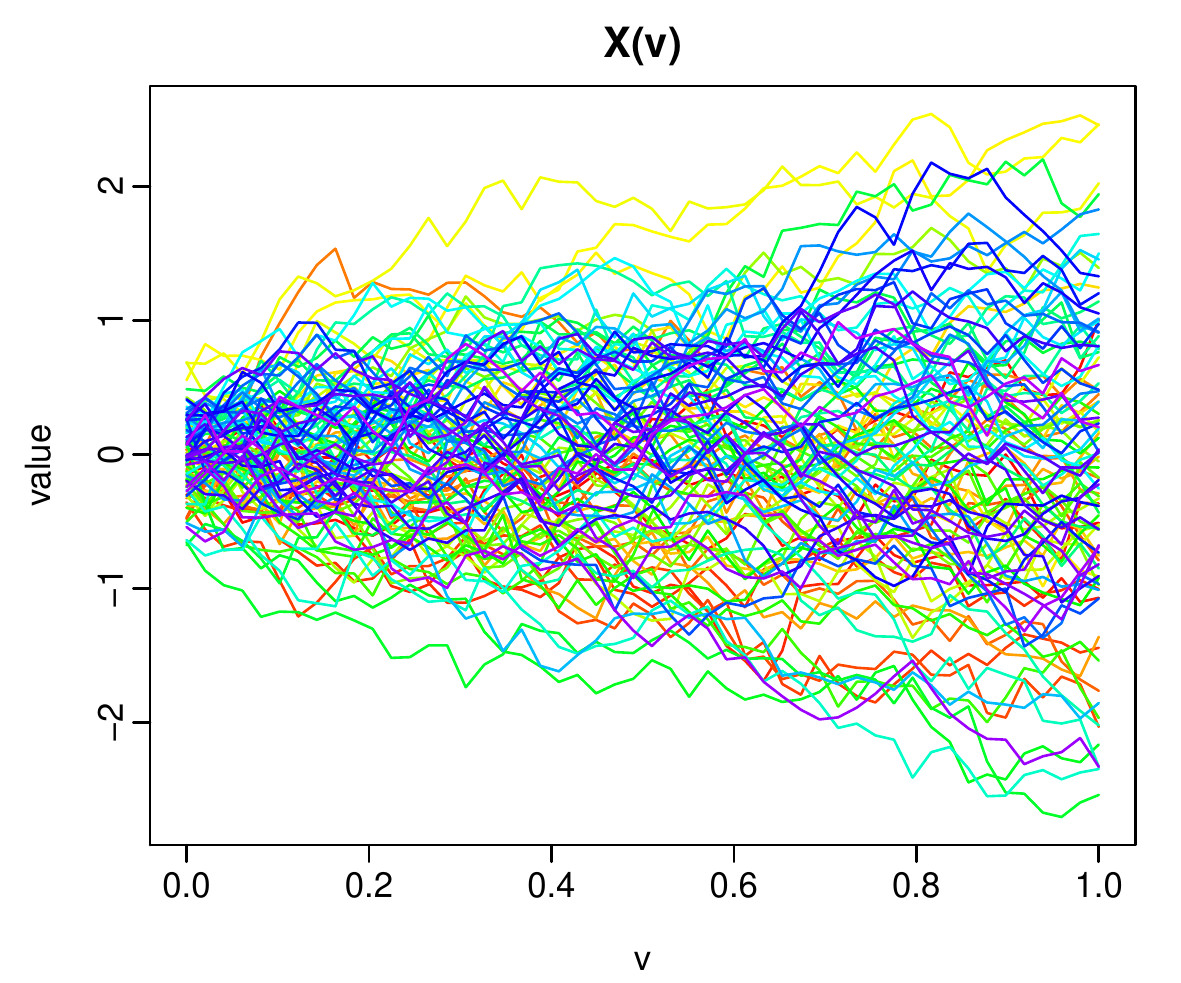}}
\caption{\small Functional time series plots of the generated functional response $\Y_1(u)$, \dots, $\Y_n(u)$, and functional predictor $\X_1(v)$, \dots, $\X_n(v)$.}\label{fig:sim_plots}
\end{figure}

A Monte-Carlo experiment under different sample sizes is conducted to present the usefulness of the proposed GcGMC. In the Monte Carlo experiment, we consider the following bivariate functional time series:
\begin{equation}
\Y_t(u) = 0.6 \Y_{t-1}(u) + \int_{v=0}^1 \X_t(v) \beta(u,v) dv + \varepsilon_t(u), \qquad u \in [0,1],\label{eq:sim}
\end{equation}
where $\varepsilon_t(u)$ denotes an independently and identically distributed (iid) error function with mean zero and finite second-order moment, $\Y_t(u)$ denotes a functional response variable and $\beta(u,v) = \sqrt{u v}$, and $\X_t(v)$ denotes a functional predictor variable that is generated from the functional autoregressive of order 1 process as follows:
\begin{equation*}
\X_t(v) = \int_0^1 \psi(v,s) \X_{t-1}(s) ds + B_t(v),
\end{equation*}
where $B_t(v)$ is a realization of a iid standard Brownian motion and $\psi(v,s) = 0.34 \exp^{\frac{1}{2} \left( v^2+s^2 \right)}$. Equation~\eqref{eq:sim} is a functional extension of the plant equation in the engineering literature \citep[see, e.g.,][Equation (2)]{Granger88}. An example of the generated bivariate functional time series is presented in Figure \ref{fig:sim_plots}.

Throughout the experiments, $n = [250, 500, 1000]$ functions are generated at $p$ equally spaced points in the interval $u, v \in [0,1]$. For each sample size, the generated data are divided into the training and test samples with sizes $0.8\times n$ and $0.2 \times n$, respectively. Using the first $0.8 \times n$ of the data, we obtained one-step-ahead forecasts of $\X_t(v)$ and $\Y_t(u)$ at time $t = 0.8\times n + 1$. Having increased the training sample by one, we then obtain-one step-ahead forecasts of $\X_t(v)$ and $\Y_t(u)$ at time $t = 0.8\times n + 2$. This process is repeated until the training sample covers the entire data. For each sample size, we repeat this procedure 100 times and for each time, we compute the GcGMC values of the predictor and response variables. If GcGMC($\Y$) $>$ GcGMC($\X$), we conclude $\Y$ is more predictable than $\X$. Further, if GcGMC($\Y)>0$ and GcGMC($\X)<0$, we conclude that $\X$ Granger-causes $\Y$, and thus $\X$ is more adequate to be the predictor. In Table~\ref{tab:simulation}, we report the number of times, where GcGMC($\Y$) $>$ GcGMC($\X$) and GcGMC($\Y)>0$, GcGMC($\X)<0$. As sample size $n$ increases from 250 to 1000, the probability of making a correct decision increases. 

\begin{table}[!htbp]
\centering
\caption{\small Out of 100 replications, we compute the number of times where $\Y$ is more predictable than $\X$ when GcGMC($\Y$) $>$ GcGMC($\X$) and $\X$ Granger-causes $\Y$ when GcGMC($\Y)>0$ and GcGMC($\X)<0$.}\label{tab:simulation}
\tabcolsep 0.36in
\begin{small}
\begin{tabular}{@{}lcccccc@{}}
\toprule
 & \multicolumn{3}{c}{GcGMC($\Y$) $>$ GcGMC($\X$)} & \multicolumn{3}{c}{GcGMC($\Y)>0$, GcGMC($\X)<0$} \\
$p\backslash n$ & 250 & 500 & 1000  & 250 & 500 & 1000  \\
\midrule
50 &  88 & 94 & 99 & 57 & 84 & 92 \\
100 & 90 & 99 & 100 & 62 & 82 & 96 \\
200 & 91 & 97 & 99 & 74 & 81 & 90 \\
400 & 96 & 99 & 99 & 71 & 85 & 92 \\
\bottomrule
\end{tabular}
\end{small}
\end{table}

\section{Data analyses}\label{sec:4}

\subsection{Sea surface temperature and sea-level atmospheric pressure}\label{sec:31}

The Oceanic Ni\~{n}o index (ONI) is one of the primary indices used to monitor the El Ni\~{n}o Southern Oscillation. The National Oceanic and Atmospheric Administration (\url{https://www.noaa.gov}) agency considers El Ni\~{n}o conditions to be present when the ONI is +0.5 or higher, indicating the east-central tropical Pacific is significantly warmer than usual. La Ni\~{n}a conditions exist when the ONI is -0.5 or lower, indicating the region is cooler than usual. The ONI is computed by averaging 3-month sea surface temperature anomalies in an area of the east-central equatorial Pacific ocean, called the Ni\~{n}o 3.4 region (5S to 5N; 170W to 120W). The 3-month running average of sea surface temperature is often compared to a 30-year average as an indicator of the climate becoming warmer or cooler.

The Southern Oscillation Index (SOI) is a measure of the strength of the Walker circulation. It is a critical atmospheric index for controlling the strength of El Ni\~{n}o and La Ni\~{n}a events. The SOI measures the difference in sea-level atmospheric pressure between Tahiti and Darwin, Australia. By standardizing sea-level atmospheric pressures in Tahiti and Darwin, the SOI is obtained by a ratio between the difference of the two standardized sea-level atmospheric pressures and their normalized standard deviation. 

\begin{figure}[!htbp]
\centering
\includegraphics[width = 8.2cm]{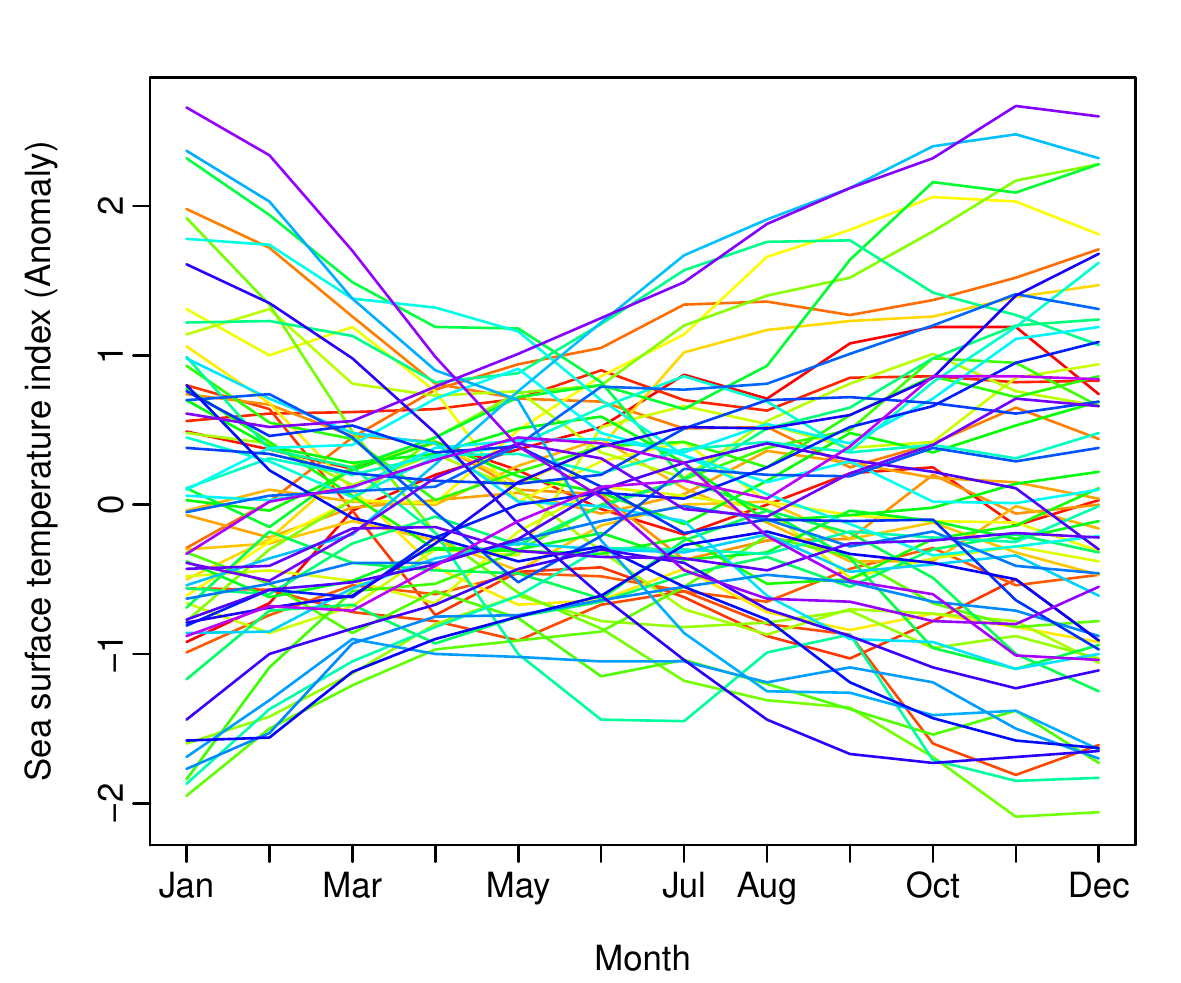}
\quad
\includegraphics[width = 8.2cm]{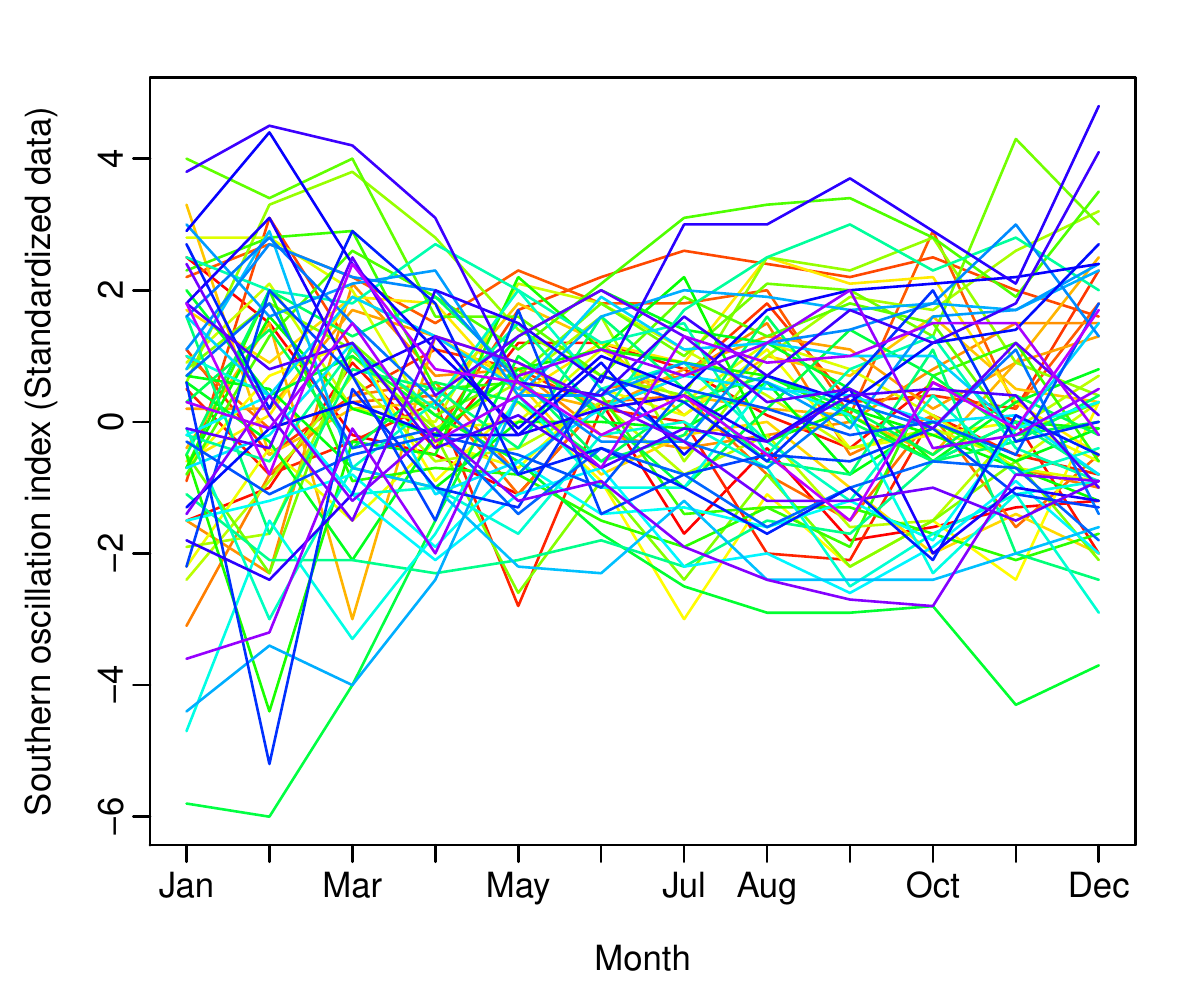}
\caption{\small Functional time series plots of the sea surface temperature anomaly and southern oscillation index (a standardized measure of sea-level atmospheric pressure) observed from January 1951 to December 2018. Each curve represents data in a year.}\label{fig:1}
\end{figure}

We consider the bivariate curve time series for the monthly sea surface temperature and sea-level atmospheric pressure from January 1951 to December 2018. The data sets can be obtained from \url{https://www.cpc.ncep.noaa.gov/data/indices/soi} and \url{https://www.cpc.ncep.noaa.gov/products/analysis_monitoring/ensostuff/detrend.nino34.ascii.txt}. In Figure~\ref{fig:1}, we plot the two functional time series. 

We implement a functional nonparametric regression to predict the sea-level atmospheric pressure and sea surface temperature. We split our entire sample into an initial training sample consisting of years from 1951 to 1983 and a testing sample composed of years from 1984 to 2018. Using the years from 1951 to 1983, we obtain one-step-ahead forecasts of the sea-level atmospheric pressure and sea surface temperature in 1984. Then, we increase the training sample by one year and obtain one-step-ahead forecasts of the sea-level atmospheric pressure and sea surface temperature in 1985, respectively. We obtain our forecasts through this expanding window approach until the training sample covers the entire data sample.

Let $\X$ be the sea-level atmospheric pressure, and $\Y$ be the sea surface temperature. With the 35 years of forecasts and their corresponding forecast errors, we compute the GcGMC values of~\eqref{eq:10} and~\eqref{eq:11}, namely GcGMC$(\X) = 0.0093$ and GcGMC$(\Y) = -0.2441$. Given $\text{GcGMC}(\X) > \text{GcGMC}(\Y)$, we conclude $\X$ is more predictable than $\Y$. Because the GcGMC of the sea-level atmospheric pressure is greater than 0, the inclusion of the sea surface temperature can help to reduce the sum squared prediction errors in the prediction of the sea-level atmospheric pressure. Because the GcGMC of the sea surface temperature is less than 0, the inclusion of the sea-level atmospheric pressure cannot help to reduce the sum squared prediction errors in the prediction of the sea surface temperature. Therefore, we conclude that the sea surface temperature Granger-causes the sea-level atmospheric pressure.

\subsection{Dow-Jones Industrial Average and its constituent stocks}\label{sec:32}

The Dow-Jones Industrial Average (DJIA) is a stock market index that shows how 30 large publicly owned companies based in the United States have traded during a standard New York Stock Exchange trading session. We consider daily cross-sectional returns from 2/January/2018 to 31/December/2018. The data were obtained from the Thompson Reuters DataScope Tick History. We have a sample of log-price observations, denoted by $q_t^j(u_i)$ for each day $t=1,\dots,n$. We define the $j$\textsuperscript{th} log return on day $t$ as
\begin{equation*}
r^j_t(u_i) = q_t^j(u_{i+1}) - q_t^j(u_i), \qquad i=1,\dots,95, 
\end{equation*}
that is, $r^j_t(u_i)$ is the log return for $j$\textsuperscript{th} company at the middle of time interval $i$ at day $t$ \citep[e.g., see][]{KRS17, SYK19, LRS20}. In a given day, there are 96 5-minute price observations from 9:30 to 17:20 Eastern time, resulting in 95 values of the log returns. 

In Table~\ref{tab:DJIA_stocks}, we list 30 constituent stocks of the DJIA. There are five stocks, namely Cisco, Dow Chemical, Intel, Microsoft, and Walgreen, that have missing observations, and are thus removed from our analysis for consistency.

\begin{table}[!htbp]
\centering
\tabcolsep 0.03in
\caption{\small Stock name and tick symbol of 30 constituent of the Dow-Jones index}\label{tab:DJIA_stocks}
\begin{small}
\begin{tabular}{@{}llllll@{}}
\toprule
Stock & Tick symbol & Stock & Tick symbol & Stock & Tick symbol \\
\midrule
Apple Inc & AAPL.OQ  & IBM & IBM & PFE & Pfizer \\
AXP & American Express & INTC.OQ & Intel & PG & Procter \& Gamble \\
BA & Boeing & JNJ & Johnson \& Johnson & TRV & Travelers Companies Inc \\
CAT & Caterpillar & JPM & JPMorgan Chase & UNH & United Health \\
CSCO.OQ & Cisco & KO & Coca-Cola & UTX & United Technologies \\
CVX & Chevron & MCD & McDonald's & V & Visa \\
DIS & Disney & MMM & 3M & VZ & Verizon \\
DOW & Dow Chemical & MRK & Merck & WBA.OQ & Walgreen \\
GS & Goldman Sachs & MSFT.OQ & Microsoft & WMT & Wal-Mart \\
HD & Home Depot & NKE & Nike & XOM & Exxon Mobil \\
\bottomrule
\end{tabular}
\end{small}
\end{table}

We implement a functional nonparametric regression to predict the DJIA, and each of the 25 stocks. We split our entire sample into an initial training sample consisting of days from 2/January/2018 to 7/August/2018 (151 days in total) and a testing sample composed of days 8/August/2018 to 31/December/2018 (100 days in total). Using days from 2/January/2018 to 7/August/2018, we obtain one-step-ahead forecasts of the DJIA and each of its constituent stocks on day 8/August/2018. Then, we increase the training sample by one day and obtain one-step-ahead forecasts of the DJIA and each of its constituent stock in day 9/August/2008, respectively. Through this expanding-window approach, we obtain our forecasts until the training sample covers the entire data sample.

With 100 days of forecasts and their corresponding forecast errors, we compute the GcGMC values when either a stock or DJIA is a response variable in the nonparametric function-on-function regression in Table~\ref{tab:GcGMC_DJIA}. Let $\X$ be the log-return of DJIA and $\Y$ be the log-return of a stock. For several stocks highlighted in blue in the table, we observe that the GcGMC of the DJIA is less than 0, and the GcGMC of the stock is greater than 0. Thus, the DJIA's inclusion can help reduce the sum squared prediction errors in the stock's prediction. These stocks highlighted in blue are lagging behind the DJIA for the period we considered. When the two GcGMC values have the same sign, we can no longer make a specific statement about leading and lagging. Instead, we can only highlight those stocks that are less predictive than the DJIA in red as GcGMC$(\X)$ $>$ GcGMC$(\Y)$; similarly, those stocks that are more predictive than the DJIA in white. For those stocks that are lagging behind the DJIA, a possible practical implication is to predict the direction of their log returns based on the most recent log return of the DJIA.

\begin{table}[!htbp]
\tabcolsep 0.045in
\centering
\caption{\small GcGMC values when either the log return of a stock or DJIA is the response variable in the nonparametric function-on-function regression.}\label{tab:GcGMC_DJIA}
\begin{small}
\begin{tabular}{@{}lrlrlrlrlr@{}}
\toprule
Variable & GcGMC & Variable & GcGMC & Variable & GcGMC & Variable & GcGMC & Variable & GcGMC  \\
\midrule
AAPL.OQ & \cellcolor{blue!25}{0.0031} & AXP &  \cellcolor{blue!25}{0.0058} & JNJ &  \cellcolor{blue!25}{0.0039} & KO & \cellcolor{blue!25}{0.0006}  & MRK & \cellcolor{blue!25}{0.0032} \\
DJIA & \cellcolor{blue!25}{-0.0185} & DJIA &  \cellcolor{blue!25}{-0.0063} & DJIA &  \cellcolor{blue!25}{-0.0064} & DJIA & \cellcolor{blue!25}{-0.0055}  & DJIA & \cellcolor{blue!25}{-0.0201} \\
\\
NKE & \cellcolor{blue!25}{0.0015}  & PFE & \cellcolor{blue!25}{0.0013} & UNH & \cellcolor{blue!25}{0.0019} &  UTX & \cellcolor{blue!25}{0.0031} & WMT & \cellcolor{blue!25}{0.0007} \\
DJIA & \cellcolor{blue!25}{-0.0133}  & DJIA & \cellcolor{blue!25}{-0.0140} & DJIA & \cellcolor{blue!25}{-0.0087} &  DJIA & \cellcolor{blue!25}{-0.0127} & DJIA & \cellcolor{blue!25}{-0.0061} \\
\\
TRV & \cellcolor{red!25}{-0.6959} & BA & -0.0022 & CAT & -0.0040 & CVX & -0.0014 & DIS & -0.0001 \\
 DJIA & \cellcolor{red!25}{-0.0107} & DJIA & -0.0125 & DJIA & -0.0091 & DJIA & -0.0062 & DJIA & -0.0046 \\
\\
GS & -0.0034 & HD & -0.0031 & IBM & 0.0047  & JPM & -0.0046 & MCD & -0.0003 \\
DJIA & -0.0121 & DJIA & -0.0073 & DJIA & 0.0141  & DJIA & -0.0052 & DJIA & -0.0107 \\
\\
MMM & -0.0015 & PG & -0.0071 & V & -0.0011  & VZ & -0.0051 & XOM & -0.0004 \\
DJIA & -0.0102 & DJIA & -0.0079 & DJIA & -0.0024 & DJIA & -0.0132 & DJIA & -0.0074 \\
\bottomrule
\end{tabular}
\end{small}
\end{table}

\subsection{S\&P 500 index and WTI oil price}

The S\&P 500 index is a stock market price representing the performance of around 500 of the largest U.S. companies. When observing the S\&P 500 index, we also observe oil prices from West Texas Intermediate (WTI). We aim to investigate if the oil price is a leading or lagging variable of the S\&P 500. We collected monthly S\&P 500 indexes and WTI prices from January/1984 to December/2019 (432 months in total). For each month in a given year $t$, we compute the normalized prices of the S\&P 500 and WTI by taking into account the consumer price index in the United States (CPI). The normalized prices can be expressed as
\begin{align*}
q_t^{\text{WTI}}(u_i) &= \frac{\text{WTI}_{\text{observed}, t}(u_i)}{\text{CPI}_{\text{observed}, t}(u_i)}\times 100, \\
q_t^{\text{S\&P 500}}(u_i) &= \frac{\text{S\&P 500}_{\text{observed}, t}(u_i)}{\text{CPI}_{\text{observed}, t}(u_i)}\times 100,
\end{align*}
where $i$ denotes any given month. The log returns of the normalized WTI and S\&P 500 at month $i$ in year $t$ can be expressed as
\begin{align*}
r_t^{\text{WTI}}(u_i) &= \ln \left[\frac{q_t^{\text{WTI}}(u_{i+1})}{q_t^{\text{WTI}}(u_i)}\right], \\
r_t^{\text{S\&P 500}}(u_i) &= \ln \left[\frac{q_t^{\text{S\&P 500}}(u_{i+1})}{q_t^{\text{S\&P 500}}(u_i)}\right],   \qquad i=1,\dots,11.
\end{align*}

\begin{figure}[!htbp]
\centering
{\includegraphics[width=8.2cm]{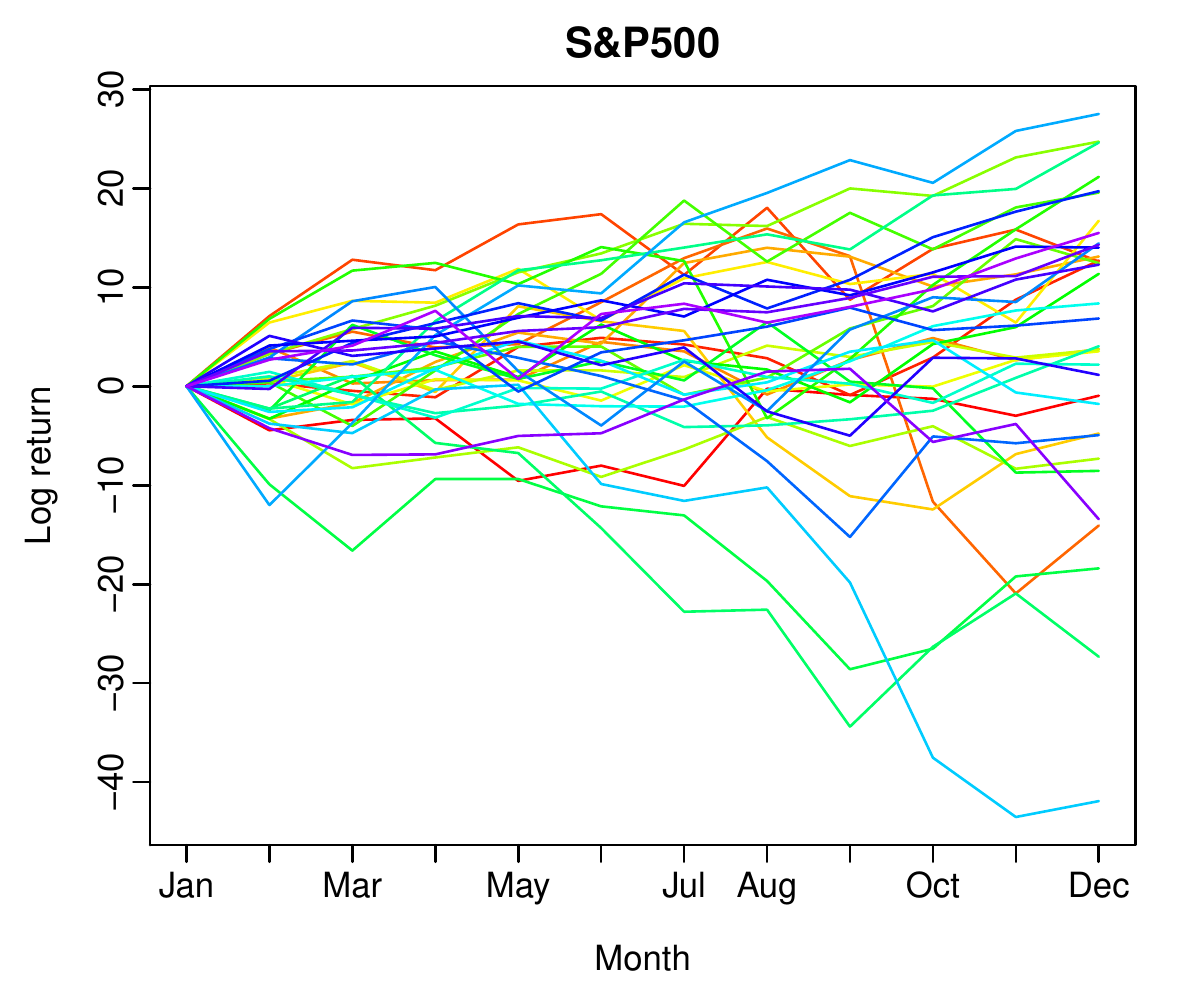}}
\quad
{\includegraphics[width=8.2cm]{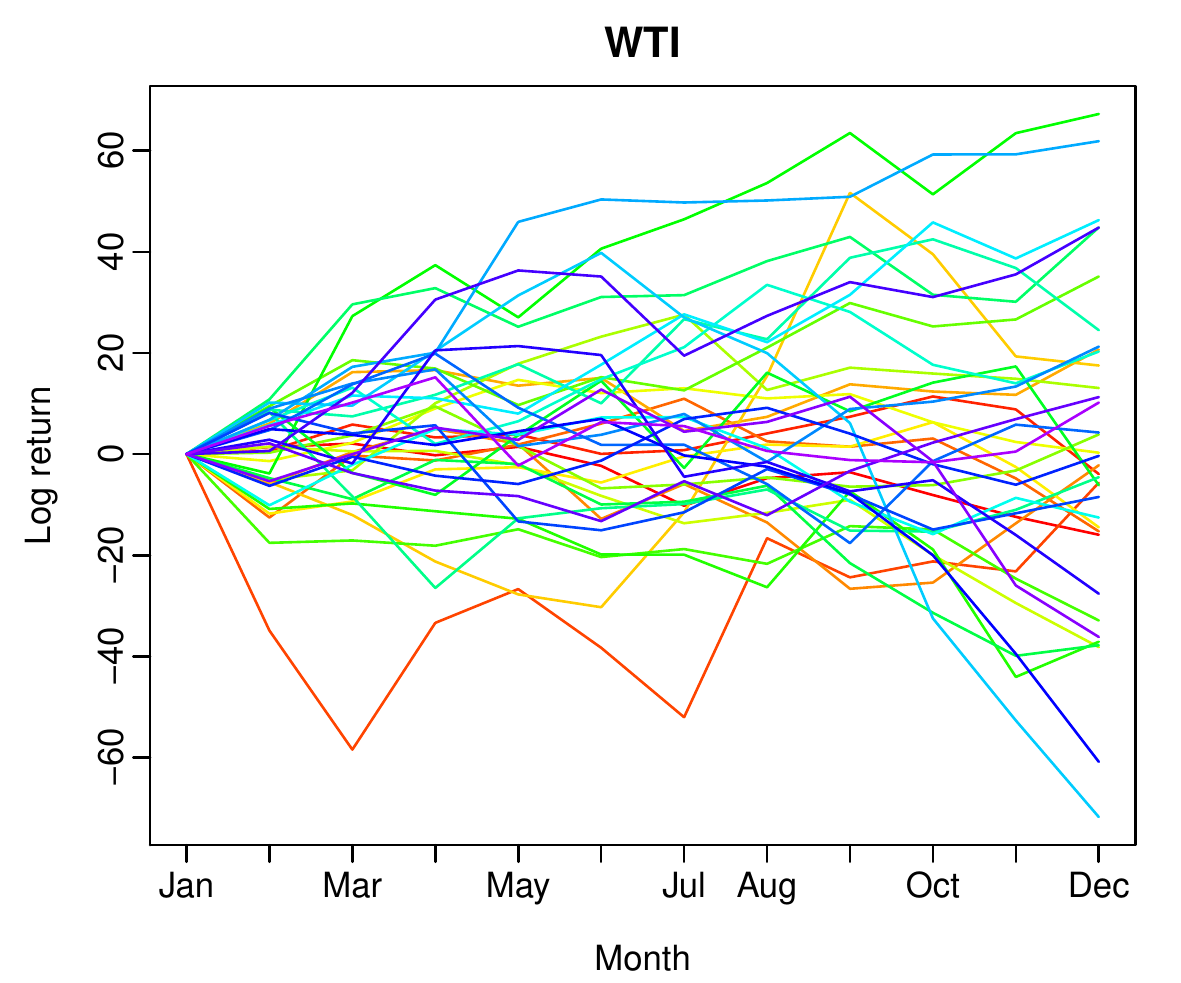}}
\caption{\small Functional time series plots of the log returns of the normalized prices of the S\&P 500 and WTI by incorporating the consumer price index in the United States.}\label{fig:sp_wti}
\end{figure}

In Figure~\ref{fig:sp_wti}, we display two functional time series plots of the log returns of the S\&P 500 and WTI. Using a stationarity test of \cite{HKR14}, we checked and verified that both series are stationary with $p$-values of 0.998 and 0.931, respectively.

We implement a nonparametric function-on-function regression to predict the log returns of S\&P 500 and WTI. The entire data set was split into an initial training sample consisting of monthly log returns from January/1984 to December/2003 (20 curves in total) and a testing sample composed of months from January/2004 to December/2019 (16 curves in total). Using the initial training sample, we obtain one-step-ahead curve forecasts of the log returns of the WTI and S\&P 500 in 2004. Then, we increase the number of curves in the training sample by one year and obtain one-step-ahead forecasts of the WTI and S\&P 500 in 2005. Through an expanding-window approach, we obtain our forecasts until the training sample covers the entire data sample.

Let $\X$ be the S\&P 500 log returns and $\Y$ be the WTI log returns. With 16 curves in the testing sample, we compute the corresponding one-step-ahead forecast errors, from which we compute the GcGMC when the S\&P 500 or WTI is the response variable, namely GcGMC$(\X) = -0.0232$ and GcGMC$(\Y) = 0.1255$. We observe that the GcGMC of the WTI is greater than 0, and the GcGMC of the S\&P 500 is less than 0, so the inclusion of the S\&P 500 can help reduce the sum squared prediction errors in the prediction of the WTI. Thus, the S\&P 500 index is a leading variable of the WTI price. A possible practical implication is to predict the direction of the WTI price based on the most recent S\&P 500 index.

\section{Conclusion}\label{sec:5}

We extend the Granger causality generalized measures of correlation from bivariate scalar to curve time series. With this measure, we can investigate which curve time series Granger-causes the other one; in turn, it helps to determine suitable predictor and response variables. Granger causality can be viewed from a prediction viewpoint. The measure can be computed by a nonparametric function-on-function regression that captures the possible nonlinear pattern between the predictor and response. Illustrated by a climatology data set, we find that the sea surface temperature Granger-causes the sea-level atmospheric pressure. From the Dow-Jones index data set, we find those constituent stocks that are lagged behind the Dow-Jones index. From the S\&P 500 and WTI oil data sets, we find that the S\&P 500 index Granger causes the WTI price.

There are two limitations associated with our Granger causality generalized measures of correlation. This measure depends on the length of training and testing samples. Sometimes, an outlying observation in the testing sample can affect the estimation accuracy of our proposed measure. Second, we compute a one-step-ahead forecast and its errors as a way of assessing predictive ability. While this sets up the foundation for our measure, it is sometimes more useful to consider other longer-term forecast horizons. 

There are several ways in which the current work may be further extended, and we briefly outline three. First, we could extend Granger causality tests studied in \cite{Geweke84} to function-valued variables, and possibly assess the strength of this relationship \citep[see, e.g.,][]{TBE14}. Second, since there is a strong connection between Granger causality and co-integration \citep[see, e.g.,][]{Granger86, Granger88}, one may study the concept of co-integration in non-stationary functional time series. Finally, the causality arisen from the Granger causality generalised measures of correlation may depend on a specific forecast horizon. We may extend the current work by forecasting $h>1$ step ahead, which may help to incorporate seasonality in a functional time series forecasting \citep[see, e.g.,][]{CMZ19}.



\bibliographystyle{agsm}
\bibliography{GcGMC_FTS.bib}

\end{document}